\documentclass[useAMS,usenatbib]{mnras}
\usepackage{graphics,epsfig,psfig}
\usepackage[normalem]{ulem}
\usepackage{xcolor}
\usepackage{float}
\usepackage{subfig}
\usepackage{graphicx}
\usepackage[]{inputenc,amssymb}
\usepackage{gensymb}
\usepackage{todonotes}
\usepackage{bm}
\usepackage{ulem}
\usepackage{latexsym}
\usepackage{epsfig}

\def\aap{A\&A}
\def\apj{ApJ}
\def\apjs{ApJS}
\def\apjl{ApJL}
\def\mnras{MNRAS}

\def\nat{Nature}

\def\araa{ARA\&A}       
\def\pasp{PASP}              

\renewcommand{\vec}[1]{\bm{#1}}

\title[Hubble parameter using GW190521]{First measurement of the Hubble parameter from bright binary black hole GW190521}
\author[Mukherjee et al. (2020)]{Suvodip Mukherjee$^{1}$\thanks{ s.mukherjee@uva.nl, mukherje@iap.fr}\thanks{The author list is in the alphabetical order except the corresponding author.}, Archisman Ghosh$^{1,2,3,4}$, Matthew J. Graham$^{5}$,  \newauthor
Christos Karathanasis$^{6}$,
Mansi M. Kasliwal$^{5}$,
Ignacio Maga\~na Hernandez$^{7}$, \newauthor
Samaya M. Nissanke$^{1}$,  Alessandra Silvestri$^{2,3}$, 
Benjamin D. Wandelt$^{8,9,10}$\\
\\
$^{1}$ Gravitation Astroparticle Physics Amsterdam (GRAPPA),
Anton Pannoekoek Institute for Astronomy and Institute for Physics,\\
University of Amsterdam, Science Park 904, 1090 GL Amsterdam, The Netherlands\\
$^2$ Institute Lorentz, Leiden University, PO Box 9506, Leiden 2300 RA, The Netherlands\\
$^3$ Delta Institute for Theoretical Physics, Science Park 904, 1090 GL Amsterdam, The Netherlands\\
$^4$ Ghent University, Proeftuinstraat 86, 9000 Gent, Belgium \\
$^5$ Division of Physics, Math and Astronomy, California Institute of Technology, 1200 E. California Boulevard, Pasadena, California 91125, USA\\
$^6$ Institut de F\'{\i}sica d'Altes Energies (IFAE), Barcelona Institute of 
Science and Technology, Barcelona, Spain\\
$^7$ Department of Physics, University of Wisconsin-Milwaukee, Milwaukee, WI 53201, USA\\
$^8$ Institut d'Astrophysique de Paris,  98bis Boulevard Arago, 75014 Paris, France\\
$^9$ Sorbonne Universites, Institut Lagrange de Paris,  98 bis Boulevard Arago, 75014 Paris, France\\
$^{10}$ Center for Computational Astrophysics, Flatiron Institute, 162 5th Avenue, 10010, New York, NY, USA\\
}
\begin{document}
\label{firstpage}
\pagerange{\pageref{firstpage}--\pageref{lastpage}}
\maketitle


\begin{abstract}
The Zwicky Transient Facility (ZTF) reported the event ``ZTF19abanrhr'' as a candidate electromagnetic (EM) counterpart at a redshift $z=0.438$ to the gravitational wave (GW) emission from the binary black hole merger GW190521. Assuming that ZTF19abanrhr is the {\it bona fide} EM counterpart to GW190521, and using the GW luminosity distance estimate from three different waveforms NRSur7dq4, SEOBNRv4PHM, and IMRPhenomPv3HM, we report a measurement of the Hubble constant $H_0= 50.4_{-19.5}^{+28.1}$ km/s/Mpc, $ 62.2_{-19.7}^{+29.5}$ km/s/Mpc, and  $ 43.1_{-11.4}^{+24.6}$ km/s/Mpc (median along with $68\%$ credible interval) respectively after marginalizing over matter density $\Omega_m$ (or dark energy equation of state $w_0$) assuming the flat LCDM (or wCDM) model. Combining our results with the binary neutron star event GW170817 with its redshift measurement alone, as well as with its inclination angle inferred from Very Large Baseline Interferometry (VLBI), we find $H_0= 67.6_{-4.2}^{+4.3}$ km/s/Mpc,  $\Omega_m= 0.47_{-0.27}^{+0.34}$, and $w_0= -1.17_{-0.57}^{+0.68}$ (median along with $68\%$ credible interval) providing the most stringent measurement on $H_0$ and the first constraints on $\Omega_m$ and $w_0$ from bright standard siren. In the future, $1.3\%$ measurement of $H_0=68$ km/s/Mpc and $28\%$ measurement of $w_0=-1$ is possible from about $200$ GW190521-like sources.  
\end{abstract}
\vspace{-0.8cm}
\section{Introduction}
Gravitational waves (GWs) from mergers of compact binaries such as neutron stars or black holes have the exquisite property that they give a direct measurement of the luminosity distance to these sources  -- they are termed as \textit{standard sirens} \citep{1986Natur.323..310S, Holz:2005df, Dalal:2006qt, 2010ApJ...725..496N}. With additional information on the sources' redshift $z$, one can then use the distance-redshift relation  
$d_l^{GW}= c(1+z)\int_0^z \frac{dz'}{H(z)}$ 
to measure the cosmological parameters, in particular related to the expansion history $H(z)$, 
 such as the Hubble constant, $H_0$, the dark matter density $\Omega_m$, dark energy density $\Omega_{de}$, as well as the equation of state (EoS) of dark energy $w(z)= w_0+ w_a(z/(1+z))$. 

Through the last decades, observations of the cosmic microwave background (CMB) \citep{Spergel:2003cb, 2011ApJS..192...18K,  Ade:2015xua, Aghanim:2018eyx}, large scale structure \citep{Anderson:2013zyy, Cuesta:2015mqa,Alam:2016hwk}, and supernovae (SNe) \citep{Perlmutter:1998np, Riess:1996pa, 2010ARA&A..48..673F}, have gradually established the flat Lambda Cold Dark Matter (LCDM) as the standard model of cosmology. While in this model, the dark energy corresponds to a cosmological constant $\Lambda$, with $w=-1$, in general it can be dynamical with a constant, $w(z)=w_0$, (wCDM model) or varying EoS. In recent years, as the different methods to measure this parameter become more and more precise, tensions have started to arise around the value of the Hubble constant $H_0$. In particular, early time probes  \citep{Ade:2015xua,Abbott:2017smn, Aghanim:2018eyx} and late time probes \citep{2009ApJ...695..287R, Riess:2019cxk, Wong:2019kwg,Freedman:2019jwv} are displaying a $4$-$5\sigma$ discrepancy in their inferred value for $H_0$  \citep{Verde:2019ivm}. 

In this context, GW standard sirens offer an exciting independent probe to measure cosmological parameters, which rely solely on the assumption that General Relativity is correct at astrophysical scales \citep{1986Natur.323..310S}. Mergers of binary neutron stars and a subset of neutron star-black hole mergers are expected to result in bright electromagnetic (EM) counterparts which can provide the redshift of the source. The binary neutron star merger GW170817 and associated ultraviolet-optical-infrared counterpart \citep{GBM:2017lvd,PhysRevX.9.011001} allowed for the identification of the host galaxy NGC4993 \citep{TheLIGOScientific:2017qsa, Abbott:2017xzu}, and provided us with the first standard siren measurement of $H_0= 70.3_{-8}^{+12}$ km/s/Mpc \citep{Abbott:2017xzu, Coulter:2017wya, Kasliwal:2017ngb}. Continued monitoring of the radio afterglow of GW170817 and VLBI measurements \citep{Mooley:2018dlz} further constrained the viewing angle of the merger and led to improved measurement of $H_0= 70.3_{-5.0}^{+5.3}$ km/s/Mpc \citep{Hotokezaka:2018dfi}. Mergers of stellar-mass binary black holes are usually not expected to have bright EM counterparts unless in significantly gaseous environments, and until recently only the ``dark siren'' statistical method has been explored to constrain $H_0$ measurement from such sources, both theoretically \citep{DelPozzo:2012zz, Chen:2017rfc, PhysRevD.93.083511, Mukherjee:2018ebj, Nair:2018ign, Gray:2019ksv, Mukherjee:2020hyn, Bera:2020jhx} and empirically   \citep{Fishbach:2018gjp,Soares-Santos:2019irc,Abbott:2019yzh,Abbott:2020khf,Palmese:2020aof}. The dark standard siren measurement from the binary black hole (BBH) merger GW170814 is $H_0=70^{+40}_{-32}$ km/s/Mpc \citep{Soares-Santos:2019irc}, and that from the recently-reported merger of a black hole with a lighter object GW190814 is $H_0 = 75^{+59}_{-13}$ km/s/Mpc \citep{Abbott:2020khf}. The current joint measurement with GW170817 along with NGC4993 (no VLBI) and the dark sirens of the first and second observing runs of Advanced LIGO-Virgo (no GW190814) is $H_0 = 68^{+14}_{-7}$ km/s/Mpc \citep{Abbott:2019yzh}.

The Zwicky Transient Facility (ZTF) recently announced a possible EM counterpart, namely an active galactic nucleus (AGN) flare \citep{PhysRevLett.124.251102}, in the same sky region at the $78\%$ spatial contour of the GW event GW190521 from the merger of binary black holes (BBHs) observed by the LIGO-Virgo detectors \citep{2015CQGra..32g4001L, Acernese_2014,PhysRevLett.123.231107, PhysRevLett.123.231108} on May 21st 2019 at GPS time 1242442967.4473 \citep{PhysRevLett.125.101102}. ZTF19abanrhr's peak luminosity occurred 50 days after the GW trigger which is consistent with predictions of a BBH merger occurring and the remnant being kicked in an AGN disk \citep{McKernan:2019hqs, PhysRevLett.123.181101}. 
The redshift from the ZTF observation together with the low-latency GW localization and distance estimates by LIGO-Virgo makes it possible to measure the expansion history  $H(z)$ from the BBH event GW190521 by exploiting the luminosity distance redshift relation. 
In this Letter, we first describe the data sets which are used for this analysis in Sec. \ref{data_d}, outline briefly our methods and then detail the results of our cosmological parameter constraints for $H_0$, matter density $\Omega_M$, and dark energy EoS  $w_0$ in Sec. \ref{method} and Sec. \ref{results} respectively. We conclude in Sec. \ref{conc} with a brief discussion of the prospects of such bright GW and EM BBH merger measurements.

\begin{figure}
\centering
\includegraphics[trim={0.cm 0.cm 0.cm 0.cm},clip,width=.4\textwidth]{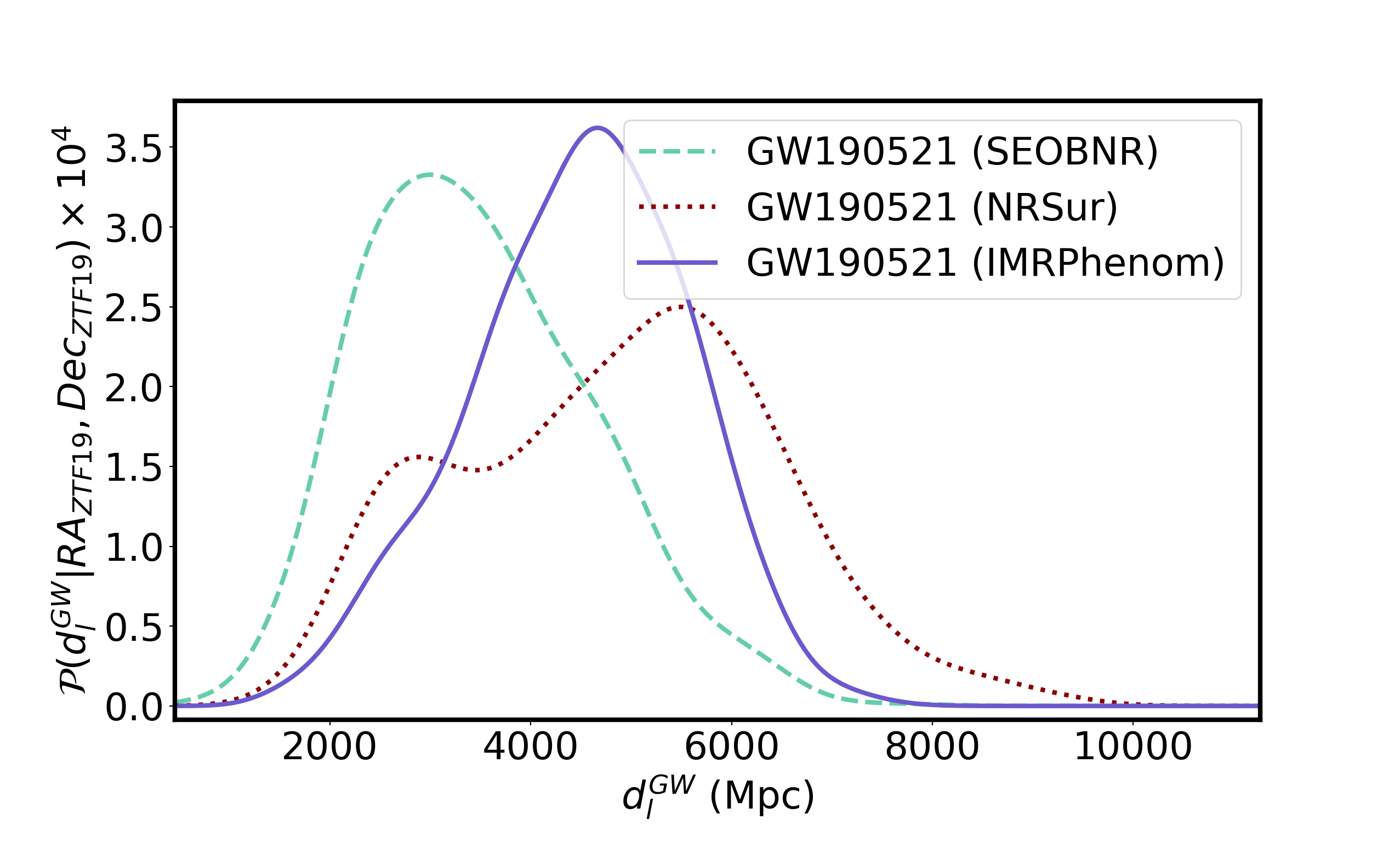}
\captionsetup{singlelinecheck=on,justification=raggedright}
\caption{Posterior of the luminosity distance along the sky direction of the EM counterpart ZTF19abanrhr to the GW source GW190521 inferred using three different waveforms (i) SEOBNRv4PHM, (ii) NRSur7dq4, (iii) IMRPhenomPv3HM.}
\label{dist}
\vspace{-0.6cm}
\end{figure}

\vspace{-0.7cm}

\section{Data products used in the  analysis}\label{data_d}

\begin{figure*}
\centering
\includegraphics[trim={0.cm 0.cm 1.cm 0.5cm},clip,width=0.73\textwidth]{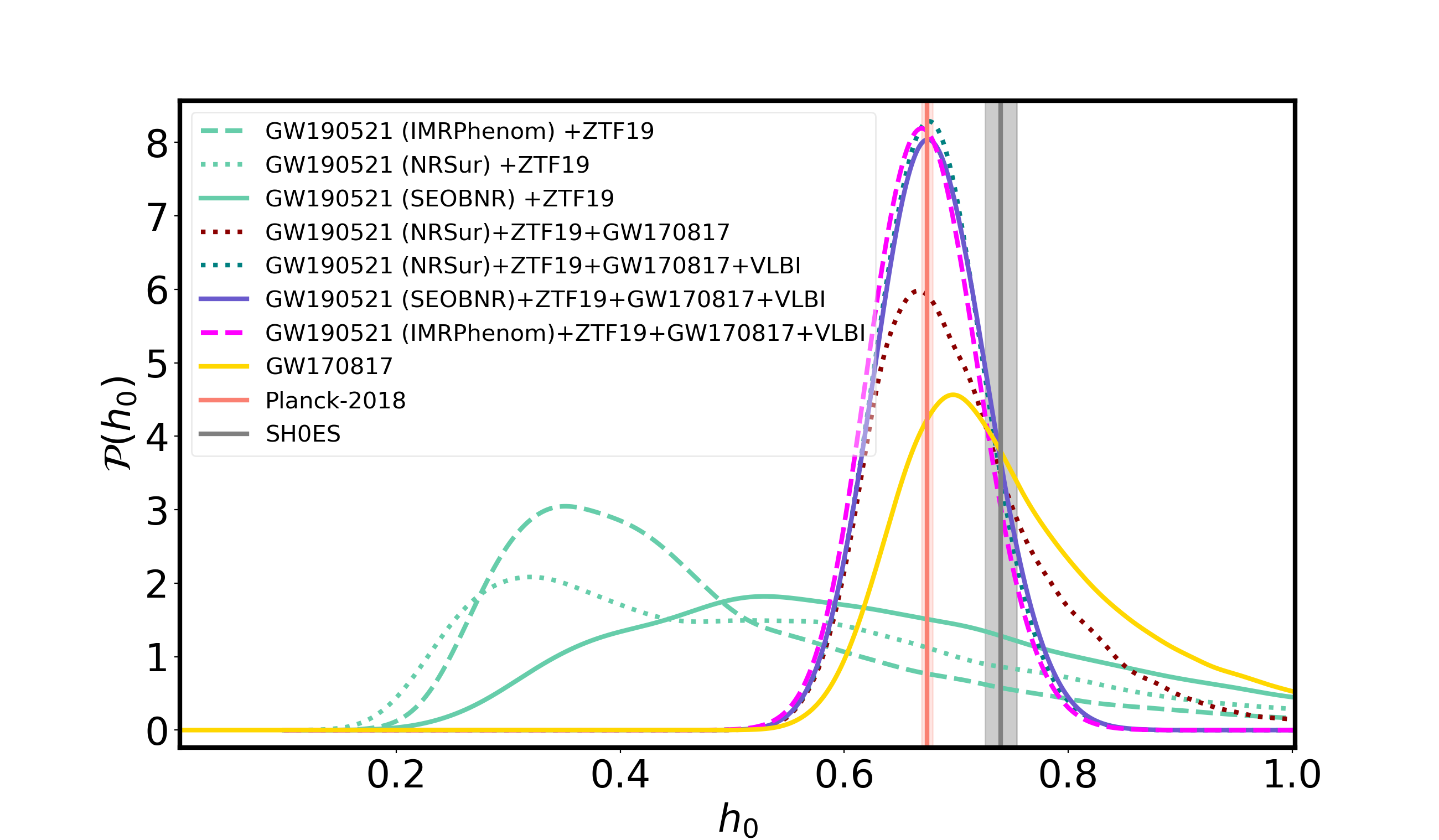}
\captionsetup{singlelinecheck=on,justification=raggedright}
\caption{Hubble constant $H_0= 100\,h_0$ km/s/Mpc estimation from GW190521 combining with the ZTF event (in green). We also show the  result on $H_0$ after including GW170817 (in red), and including the VLBI measurement of the inclination angle (in teal, blue, and magneta). For comparison we  plot the measured value of $h_0= 0.674\pm 0.005$ by the \citet{Aghanim:2018eyx} (in salmon),  SH0ES team \citep{Riess:2019cxk} $h_0=0.74 \pm 0.014$ (in grey), and GW170817 (in yellow) \citep{TheLIGOScientific:2017qsa}}
\label{allh0}
\vspace{-0.5cm}
\end{figure*}

\textit{GW190521:} The merger of two black holes each of mass $85^{+21}_{-14}M_\odot$ and $66^{+17}_{-18}M_\odot$ was detected by the Advanced LIGO-Virgo detector network \citep{PhysRevLett.123.231107, PhysRevLett.123.231108} with a false alarm rate 1 in 4900 years at a luminosity distance $d_l^{GW}= 5.3^{+2.4}_{-2.6}$ Gpc after marginalizing over the sky localisation  \citep{PhysRevLett.125.101102}. The inferred luminosity distance along the direction of the EM counterpart ZTF19abanrhr for analysis with three different GW waveforms (the effective-one-body model  SEOBNRv4PHM, the numerical relativity surrogate model NRSur7dq4, and the phenomenological model IMRPhenomPv3HM) are shown in Fig. \ref{dist}. We also show the posterior distribution along the ZTF direction for the source frame masses, and the inclination angle in Fig. \ref{fig:m2} obtained using these three waveforms.   
 
\textit{ZTF19abanrhr:} 
ZTF identified a candidate for an EM counterpart to GW190521 at the sky direction ($RA= 192.42625^\circ, Dec= 34.82472^\circ$), dubbed ZTF19abanrhr, which was first observed after 34 days from the GW detection. The candidate EM counterpart was identified in the sky area $78\%$ spatial contour of the GW signal GW190521. The signal is associated with an AGN J124942.3 + 344929 at redshift z = 0.438 \citep{PhysRevLett.124.251102}. 

\textit{GW170817:} On 17th August 2017, the LIGO and Virgo detectors observed a BNS merger GW170817, which was subsequently observed over the entire EM spectrum (e.g., \cite{Kasliwal:2017ngb, GBM:2017lvd}). In our analysis, we use the marginalised posterior probability density function (PDF) of $H_0$ as inferred from the BNS event GW170817 \citep{TheLIGOScientific:2017qsa, Abbott:2017xzu}, after implementing the peculiar velocity correction described in \citet{Mukherjee:2019qmm}. The value of the Hubble constant is $H_0= 68.3_{-7}^{+12}$   km/s/Mpc with $68\%$ credible intervals. 

\textit{Inclination angle from the jet of GW170817:} The Very Large Baseline Interferometry (VLBI) observations \citep{Mooley:2018dlz} and the afterglow light curve data (e.g., \citep{2018ApJ...868L..11M}) have enabled constraints of the inclination angle $i$ as $0.25$ rad $\leq i\, (\frac{d_l}{41 \rm{Mpc}}) \leq 0.45$ rad for GW170817. This, correspondingly, helps place tighter constraints on $H_0$;  a revised measurement of Hubble constant is $H_0= 70.3^{+5.3}_{-5.0}$  km/s/Mpc  \citep{Hotokezaka:2018dfi}. Implementing the peculiar velocity correction, we find the revised value as $H_0=68.3^{+ 4.6}_{-4.5}$    km/s/Mpc \citep{Mukherjee:2019qmm}. 
\vspace{-0.8cm}
\section{Methods}\label{method}
We compute the posterior distribution for the cosmological parameters $\Theta_c \in \{H_0, \Omega_m, w_0\}$ using the Bayes theorem \citep{bayes}
\begin{eqnarray}\label{posterior-1}
     \mathcal{P}(\Theta_c|{d}_l^{GW}, \vec d_{ZTF})
\propto &  \int dd_l\mathcal{P}({d}_l^{GW}| d_l, \vec d_{ZTF}, \Theta_c)) P(d_l)\Pi(\Theta_c)     
\end{eqnarray}
where $d_l^{GW}$ and $\vec d_{ZTF}$ are the GW luminosity distance data and ZTF data  respectively, $\mathcal{P}({d}_l^{GW}| d_l, \vec d_{ZTF}, \Theta_c))$  is the marginalised probability distribution  on the luminosity distance from GW190521. 
$\Pi(\Theta_c)$ and $\Pi(d_l)$ are the priors on the cosmological parameters and luminosity distance. The detail derivation of the Bayesian framework in given in the Appendix \ref{app2}. 
We consider uniform ($\Pi(H_0)= \mathcal{U}[10, 150]$, $\Pi(\Omega_m)= \mathcal{U}[0.1, 1]$, $\Pi(w_0)= \mathcal{U}[-2, -0.1]$). In this analysis, we obtain the results for two models (i) LCDM model with $\Theta_c \in\{H_0, \Omega_m\}$ (keeping $w_0=-1$ fixed), and (ii) wCDM model with $\Theta_c \in \{H_0, w_0\}$ (keeping $\Omega_m=0.315$ \citep{Aghanim:2018eyx}). The joint estimation of the cosmological parameters $\Omega_m$ (or $w_0$) and $H_0$ are important as the source is situated at high redshift. The results are obtained for three different combinations of data sets (see Sec. \ref{data_d} for the details) (D1) GW190521+ZTF19abanrhr, (D2) GW190521+ZTF19abanrhr+GW170817, (D3) GW190521+ZTF19abanrhr+GW170817+VLBI, each for three different choices of GW waveforms  (a) SEOBNRv4PHM, (b) NRSur7dq4, (c) IMRPhenomPv3HM\footnote{Hereafter we denote a particular combination of data set such as GW190521 (SEOBNRv4PHM) +ZTF19abanrhr as ``D1a''.}.  

\vspace{-0.85cm}
\section{Results}\label{results}
\textit{Constraints on Hubble constant $H_0$:} After marginalizing over $\Omega_m$, the posterior of $H_0$ for D1a, D1b, and D1c are shown in Fig. \ref{allh0}. The mild differences in the luminosity distance posteriors inferred using three different waveforms IMRPhenomPv3HM, NRSur7dq4, and SEOBNRv4PHM, leads to the observed difference in the Hubble constant posterior, as can be seen from the dashed,  dotted, and solid lines in green.  The median value of the Hubble constant for data sets D1a, D1b, and D1c with $68\%$ credible interval are $H_0=  62.2_{-19.7}^{+29.5}$ km/s/Mpc, $H_0=  50.4_{-19.5}^{+28.1}$ km/s/Mpc, and $H_0= 43.1_{-11.4}^{+24.6}$ km/s/Mpc respectively. The differences in the value of the Hubble constant from different methods are not statistically significant. After combining with the measurement from GW170817, the median value of the Hubble constant for D2b becomes $H_0= 69.1_{-5.8}^{+9.4}$ km/s/Mpc as shown by the dark-red colour in Fig. \ref{allh0} \footnote{We have chosen the waveform NRSur7dq4 than the other waveforms as it is calibrated with the numerical simulations.}. This improves the constraints in the higher values of $H_0$. Inclusion of the VLBI measurement provides the most stringent measurement from GW observations $H_0= 67.6_{-4.2}^{+4.3}$ km/s/Mpc as shown in Fig. \ref{allh0} for D3b. The results for D3a and D3c are $H_0= 67.7_{-4.2}^{+4.6}$ km/s/Mpc and $H_0= 67.1_{-4.2}^{+4.4}$ km/s/Mpc respectively which are consistent with the result from D3b.  The measurement from the data sets D3b is in agreement with the best-fit value of $H_0= 67.4 \pm 0.5$ km/s/Mpc from the Planck collaboration \citep{Ade:2015xua, Aghanim:2018eyx} and is about $1.6\sigma$ (assuming a Gaussian distribution) away from the SH0ES value of $H_0=74. \pm 1.4$ km/s/Mpc \citep{Riess:2019cxk}. 

\begin{figure}
\centering
\includegraphics[trim={0.cm 0.cm 0.cm 0.cm},clip,width=0.7\linewidth]{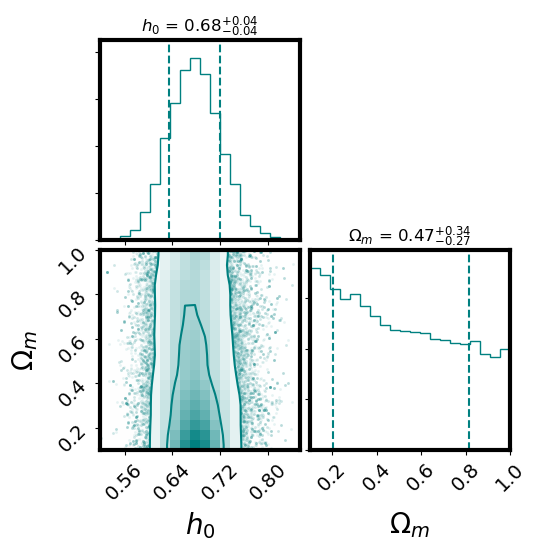}
\includegraphics[trim={0.cm 0.cm 0.cm 0.cm},clip,width=.7\linewidth]{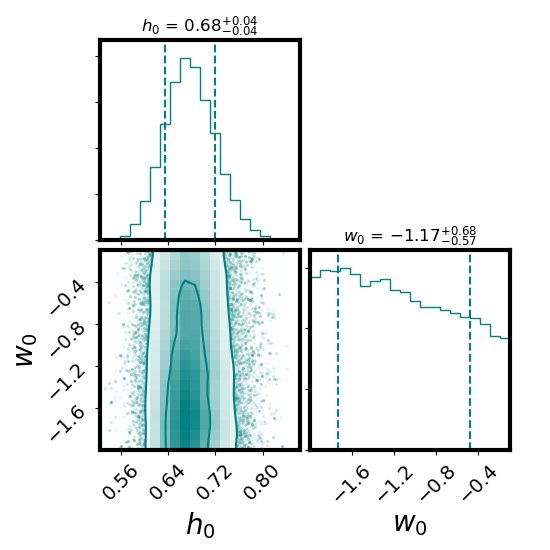}
\captionsetup{singlelinecheck=on,justification=raggedright}
\caption{The joint estimation of the Hubble constant $H_0= 100 h_0$ km/s/Mpc and (i) $\Omega_m$ with a fixed value of dark energy EoS $w_0=-1$ and (ii) $w_0$ with fixed $\Omega_m=0.315$ for the combination of data sets D3b.}
\label{oms19vlbi}
\vspace{-0.7cm}
\end{figure}

\textit{Constraints on $\Omega_m$ and $w_0$:} For the combination of datasets D3b, we show the joint parameter estimation $H_0+\Omega_m$ (for the LCDM model) and $H_0+w_0$ (for the wCDM model) in Fig. \ref{oms19vlbi}. The mean value and the $68\%$ credible interval of the matter density and dark energy EoS are  $\Omega_m= 0.47_{-0.27}^{+0.34}$ and $w_0= -1.17_{-0.57}^{+0.68}$ respectively. Though the constraints are weak, this provides the first estimation on matter density and dark energy EoS using standard sirens allowing slightly lower values. The bounds for the combination of data sets D3a and, D3c are also similar. 
With an increase in the number of GW sources, even in the absence of EM counterparts, the cosmological parameters $\Omega_m$ and $w(z)$ will also be measured accurately from the LIGO/Virgo detectors \citep{Mukherjee:2020hyn}.
\vspace{-0.9cm}
\section{Conclusion and future outlook}\label{conc}
We present here the measurement of the Hubble constant $H_0=  50.4_{-19.5}^{28.1}$ km/s/Mpc from bright standard siren GW190521 using the waveform NRSur7dq4, after marginalizing over matter density $\Omega_m$ for the LCDM model of cosmology. This is the first measurement of the Hubble constant from a BBH merger having its candidate EM counterpart detected by ZTF \citep{PhysRevLett.124.251102}. By combining the results from the BNS event GW170817 along with the constraints on the inclination angle from VLBI, we report the most stringent measurement of Hubble constant $H_0= 67.6_{-4.2}^{+4.3}$ km/s/Mpc from standard sirens. Using GW190521, we are able to obtain  constraints on the matter density $\Omega_m= 0.47_{-0.27}^{+0.34}$ and dark energy EoS $w_0= -1.17_{-0.57}^{+0.68}$ for the first time from standard sirens. Other independent analysis are also carried out adding the prior from Planck \citep{chenprep} and using GW waveforms including eccentricity \citet{gayatriprep}.

\begin{figure}
\centering
\includegraphics[trim={0.cm 0.cm 0.cm 0.cm},clip,width=0.8\linewidth]{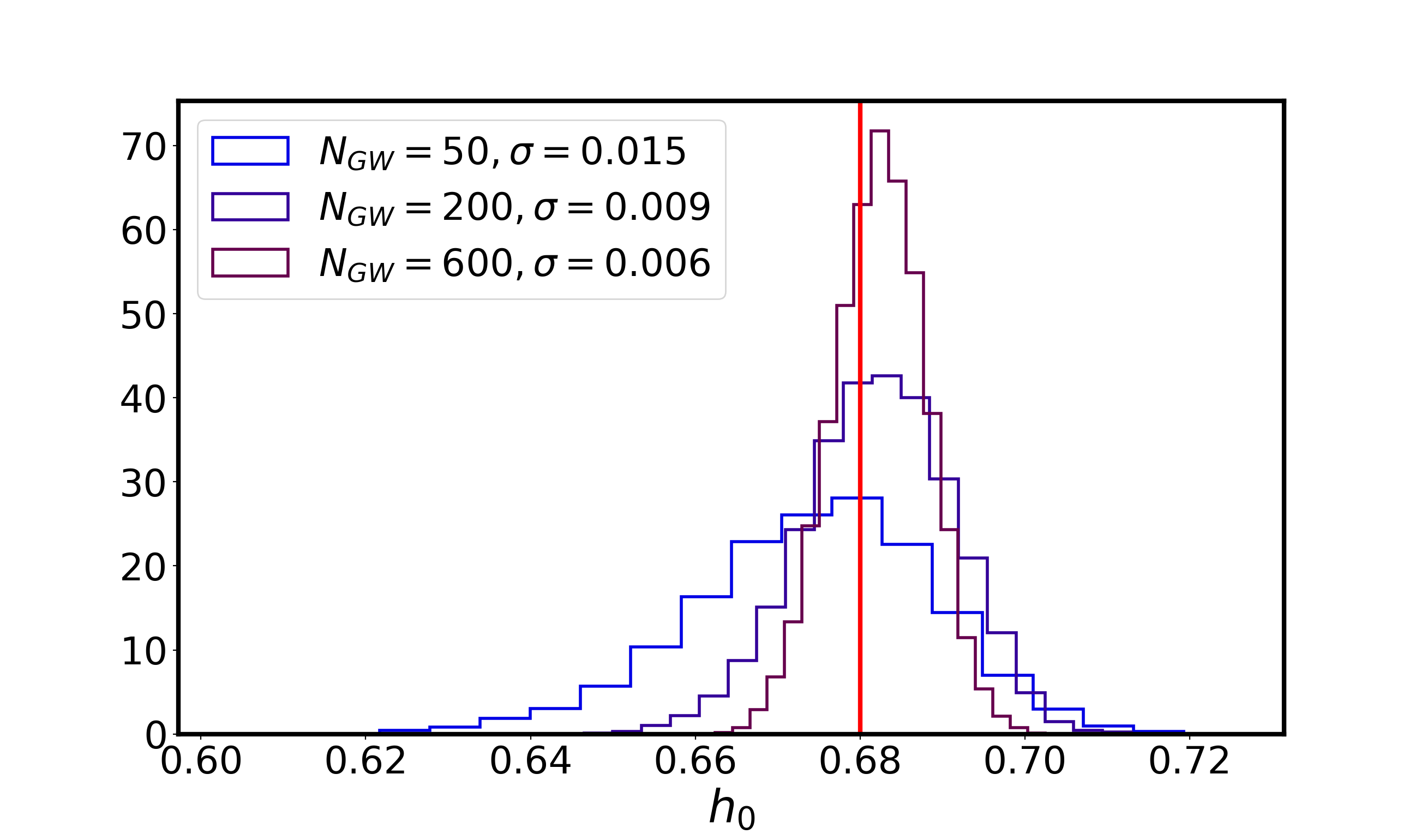}
\includegraphics[trim={0.cm 0.cm 0.cm 0.cm},clip,width=0.8\linewidth]{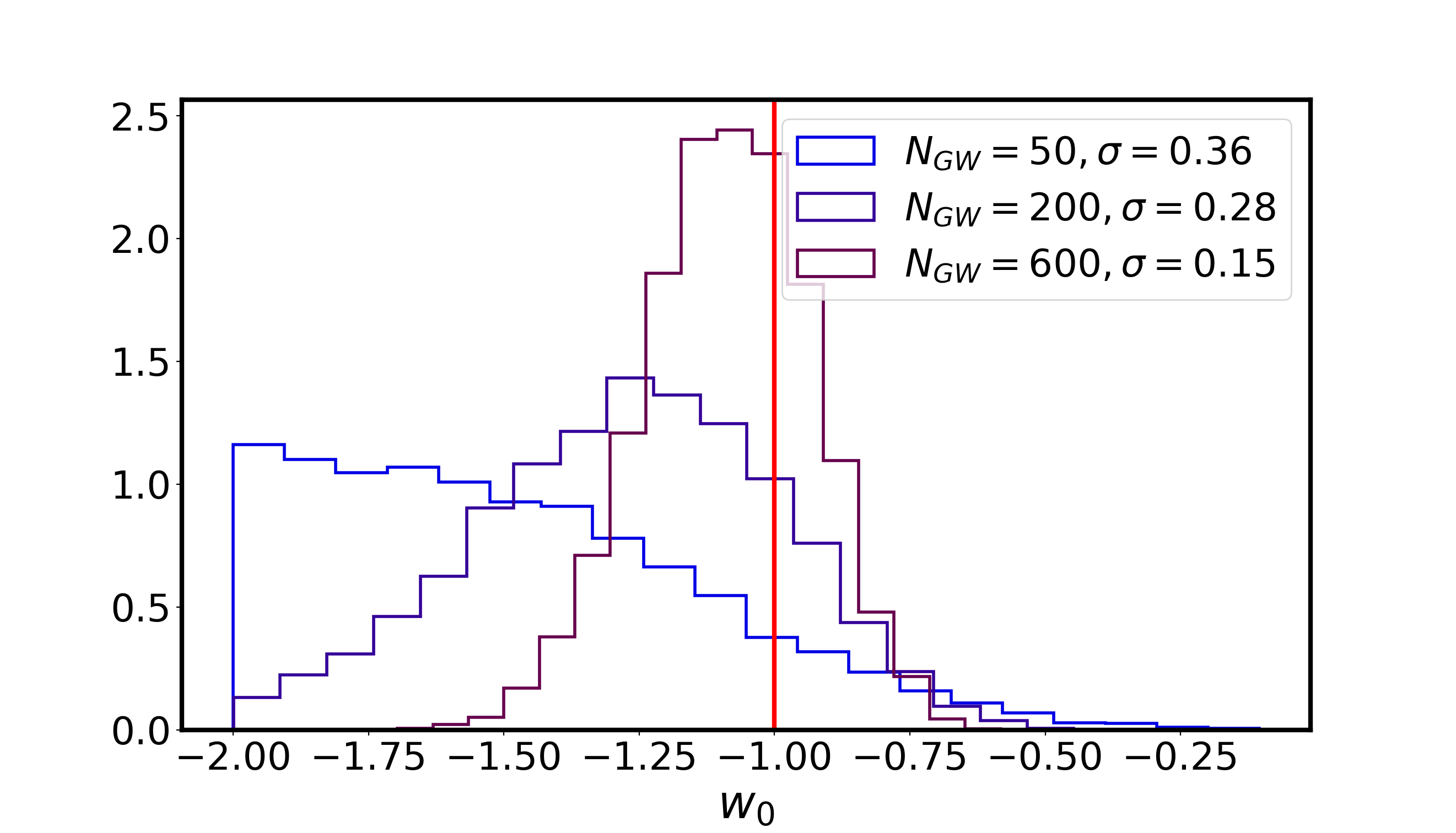}
\captionsetup{singlelinecheck=on,justification=raggedright}
\caption{Forecast: The posterior distribution on $h_0$, $w_0$ possible from GW190521-like sources detectable up to redshift $z=1$ with EM counterpart, with individual masses $85$ M$_\odot$ and $66$ M$_\odot$ in the source-frame. The line in red shows the injected value.}
\label{fig:forecast}
\vspace{-0.5cm}
\end{figure}

Future measurements with $N_{GW}$ BBHs with identified EM counterparts, can expect to beat the statistical uncertainty by $N_{GW}^{-1/2}$.  Using BBHs similar to the source-frame masses of the event GW190521 ($m_1=85$ M$_\odot$, $m_2=66$ M$_\odot$), we estimate the measurability of the cosmological parameters at LIGO/Virgo design sensitivity \citep{Acernese_2014,Martynov:2016fzi}. Considering the GW sources distributed up to redshift $z=1$ for which electromagnetic counterparts can be detected by ZTF, we show the posterior distribution on $H_0= 100 h_0$ km/s/Mpc, and $w_0$ in Fig. \ref{fig:forecast}. This shows that our method can reliably recover the injected value of the cosmological parameters with an uncertainty about $1.3\%$ on $H_0$ and with about $28\%$ on $w_0=-1$ from  $200$ GW sources.

In summary, the redshift measurement of GW190521 opens a new paradigm of measurements with multi-messenger cosmology using BBHs. Accurate identification of EM counterparts from BBHs will allow not only to measure the expansion history up to high redshift but also to explore different aspects of fundamental physics.  
This avenue is going to be useful also for the future space-based GW detector Laser Interferometer Space Antenna (LISA) \citep{2017arXiv170200786A}. LISA will detect super-massive BBHs which are also likely to have EM counterparts in gas-rich environments \citep{Armitage:2002uu, Palenzuela:2010nf, Farris:2014zjo, Haiman:2018brf}, and this observation of an EM candidate is possibly the first step towards the detection of EM counterparts on BBHs in gas-rich environments. 
\vspace{-0.85cm}
\section*{Availability of data}
The datasets were derived from sources in the public domain: \href{https://dcc.ligo.org/LIGO-P2000158/public}{https://dcc.ligo.org/LIGO-P2000158/public}. 
\vspace{-0.85cm}
\section*{Acknowledgement}
 We thank Simone Mastrogiovanni for carefully reviewing the manuscript and providing  useful suggestions. SM acknowledges useful discussions with Will M. Farr and Rachel Gray. This analysis was carried out at the Horizon cluster hosted by IAP. We thank Stephane Rouberol for smoothly running the Horizon cluster. SM and SMN are supported by the research program Innovational Research Incentives Scheme (Vernieuwingsimpuls), which is financed by the Netherlands Organization for Scientific Research through the NWO VIDI Grant No. 2016/ENW/639.042.612. AG is grateful for funding from the D-ITP. This work was supported by the GROWTH project funded by the National Science Foundation under Grant No 1545949. This work was  partially  supported   by  the Spanish MINECO   under 
the grants SEV-2016-0588 and PGC2018-101858-B-I00, some of which include 
ERDF  funds from the  European  Union. IFAE  is  partially funded by 
the CERCA program of the Generalitat de Catalunya. IMH is supported by the NSF Graduate Research Fellowship Program under grant DGE-17247915. AS acknowledges support from the NWO and the Dutch Ministry of Education, Culture and Science (OCW) (through NWO VIDI Grant No. 2019/ENW/00678104 and from the D-ITP consortium). BDW and part of the computational work are supported by the Labex ILP (reference ANR-10-LABX-63) part of the Idex SUPER,  received financial state aid managed by the Agence Nationale de la Recherche, as part of the programme Investissements d'avenir under the reference ANR-11-IDEX-0004-02.  BDW acknowledges financial support from the ANR BIG4 project, under reference ANR-16-CE23-0002. The Center for Computational Astrophysics is supported by the Simons Foundation. In this analysis, following packages are used: Corner \citep{corner}, emcee: The MCMC Hammer \citep{2013PASP..125..306F}, IPython \citep{PER-GRA:2007}, Matplotlib \citep{Hunter:2007}, NumPy \citep{2011CSE....13b..22V}, and SciPy \citep{scipy}.

\vspace{-0.9cm}
\def\urlprefix{}
\def\url#1{}
\bibliography{main.bib}
\label{lastpage}
\begin{appendix}
\section{GW source properties along the sky direction ZTF19abanrhr}
\begin{figure}
\centering
\includegraphics[trim={0.cm 0.cm 0.cm 0.cm},clip,width=.4\textwidth]{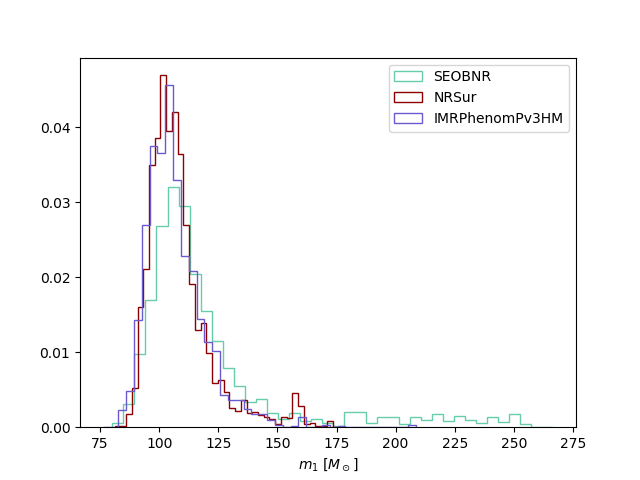}
\includegraphics[trim={0.cm 0.cm 0.cm 0.cm},clip,width=.4\textwidth]{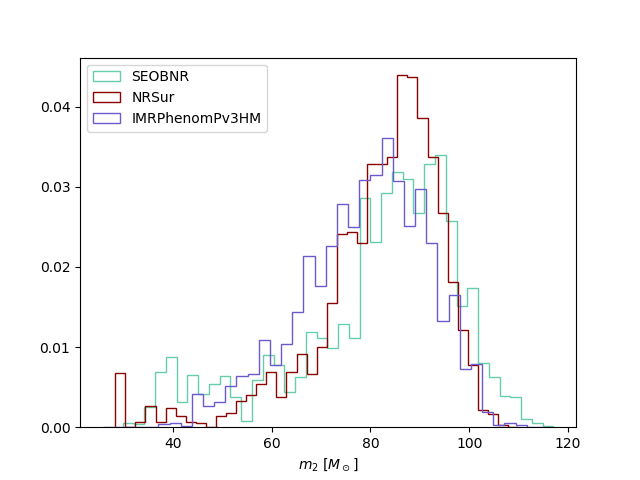}
\includegraphics[trim={0.cm 0.cm 0.cm 0.cm},clip,width=.4\textwidth]{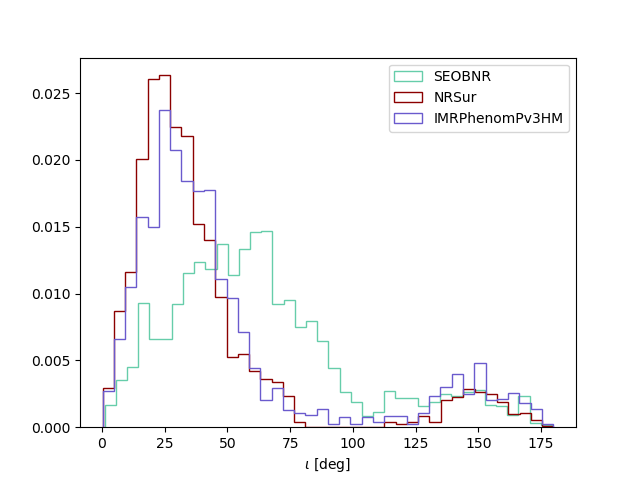}
\captionsetup{singlelinecheck=on,justification=raggedright}
\caption{We show the posterior distribution on the mass $m_1$ of the heavier component BH, $m_2$ of the lighter component BH in the source frame, and the inclination angle along the sky direction of the EM counterpart ZTF19abanrhr inferred using three different waveforms (i) SEOBNRv4PHM, (ii) NRSur7dq4, (iii) IMRPhenomPv3HM.}
\label{fig:m2}
\end{figure}

The posterior distributions for GW source masses ($m_1$ and $m_2$) and inclination angle $i$ along the line of sight of ZTF event are shown in Figs. \ref{fig:m2} for the three waveform models (i) SEOBNR, (ii) NRSur, and (iii) IMRPhenomPv3HM. {These distributions have been obtained by re-weighting the samples of the LIGO-Virgo data release with a Gaussian prior on the sky centered at the location of ZTF19abanrhr with an effective beam size of 10~sq.~deg.} The mass distributions show a bimodality for all the three  waveforms, which is most explicit for NRSur. The inclination angle shows slightly higher probability towards the values $i<90^\circ$.  

\section{Brief discussion of the Bayesian framework relevant for the analysis of GW sources with EM counterpart}\label{app2}
Let us consider a GW source detected with matched-filter signal to noise ratio $\rho >\rho_*$, where $\rho_*$ is the detection threshold. For this event, we have identified the EM counterparts within some time interval $\Delta t$ in the sky area identified from the GW observation. Using Bayes theorem \citep{bayes}, we can write the probability distribution of the EM data vector $\vec d_{EM}$ which includes its redshift $\hat z$ and sky location $(\hat \theta, \hat \phi)$ given an observed EM counterpart $\mathcal{O}_{EM}$, galaxy catalog $g$ containing the spectroscopically measured redshift, and astrophysical model $I_A$ of the source as 
\begin{eqnarray}\label{eq:em}
    &\mathcal{P}(\vec d_{EM}| d_{GW}, g, I_A)\propto \Pi(\hat z, \hat \theta, \hat \phi)\int d\Delta t \int d\Omega_{sky} \\\nonumber & \times P(g|\mathcal{O}_{EM} (t+\Delta t), z, \theta, \phi)  P(\mathcal{O}_{EM} (t+\Delta t)| d_{GW} (t), I_A) \\\nonumber & \times P(\Delta t|I_A)
\end{eqnarray}
where, $\Pi(\hat z, \hat \theta, \hat \phi)$ is the prior on the redshift of the source $\hat z$, and its sky direction $(\hat \theta, \hat \phi)$ , $P(g|\mathcal{O}_{EM} (t+\Delta t), z, \theta, \phi)$ is the likelihood of the EM data vector given galaxy catalog and EM counterpart $\mathcal{O}_{EM}$, $P(\mathcal{O}_{EM} (t+\Delta t)| d_{GW} (t), I_A)$ is the likelihood of the EM counterpart given GW data $d_{GW}$ and astrophysical model $I_A$, $P(\Delta t| I_A)$ is the probability that the signal is observed after time $\Delta t$ from GW observation for a given model $I_A$. The posterior given in Eq. \ref{eq:em} is not normalised, but this does not affect cosmological parameter inference.

The association of the GW signal and EM signal are made by observing both the signals in the time-domain. With the aid of the time-domain aspect, we can relate the redshift space information with the luminosity distance space information for a single GW source. When such association is absent, one needs to estimate the prior $P_{pop}(z, \theta, \phi|\Omega_c)$ of the allowed redshift values for a given choice of cosmological parameters $\Omega_c$
and assuming redshift distribution of the GW merger rates. If a pair of EM and GW signal is identified ($\mathcal{P}(\vec d_{EM}| d_{GW},g, I_A)$ using Eq. \ref{eq:em}), then there is no need to choose the prior $P_{pop}(z, \theta, \phi|\Omega_c)$ on the  allowed redshift and sky positions separately for a choice of cosmological parameters with an assumption on the population. One can use $\mathcal{P}(\vec d_{EM}| d_{GW},g, I_A)$ as the prior on the redshift and sky location.

\begin{figure*}
\centering
\includegraphics[trim={0.cm 0.cm 1.5cm 0.cm},clip,width=.7\textwidth]{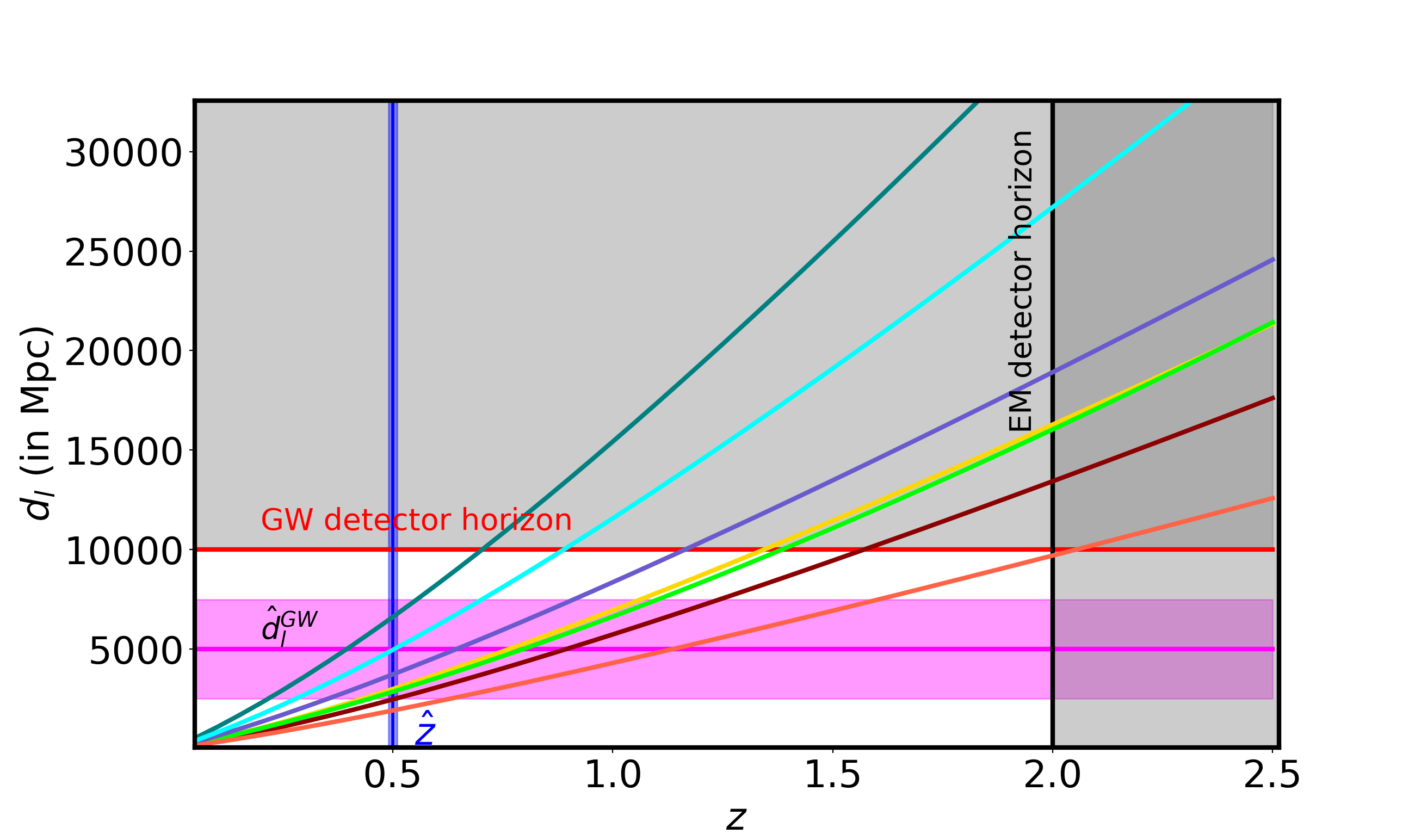}
\includegraphics[trim={0.cm 0.cm 1.5cm 1.5cm},clip,width=.7\textwidth]{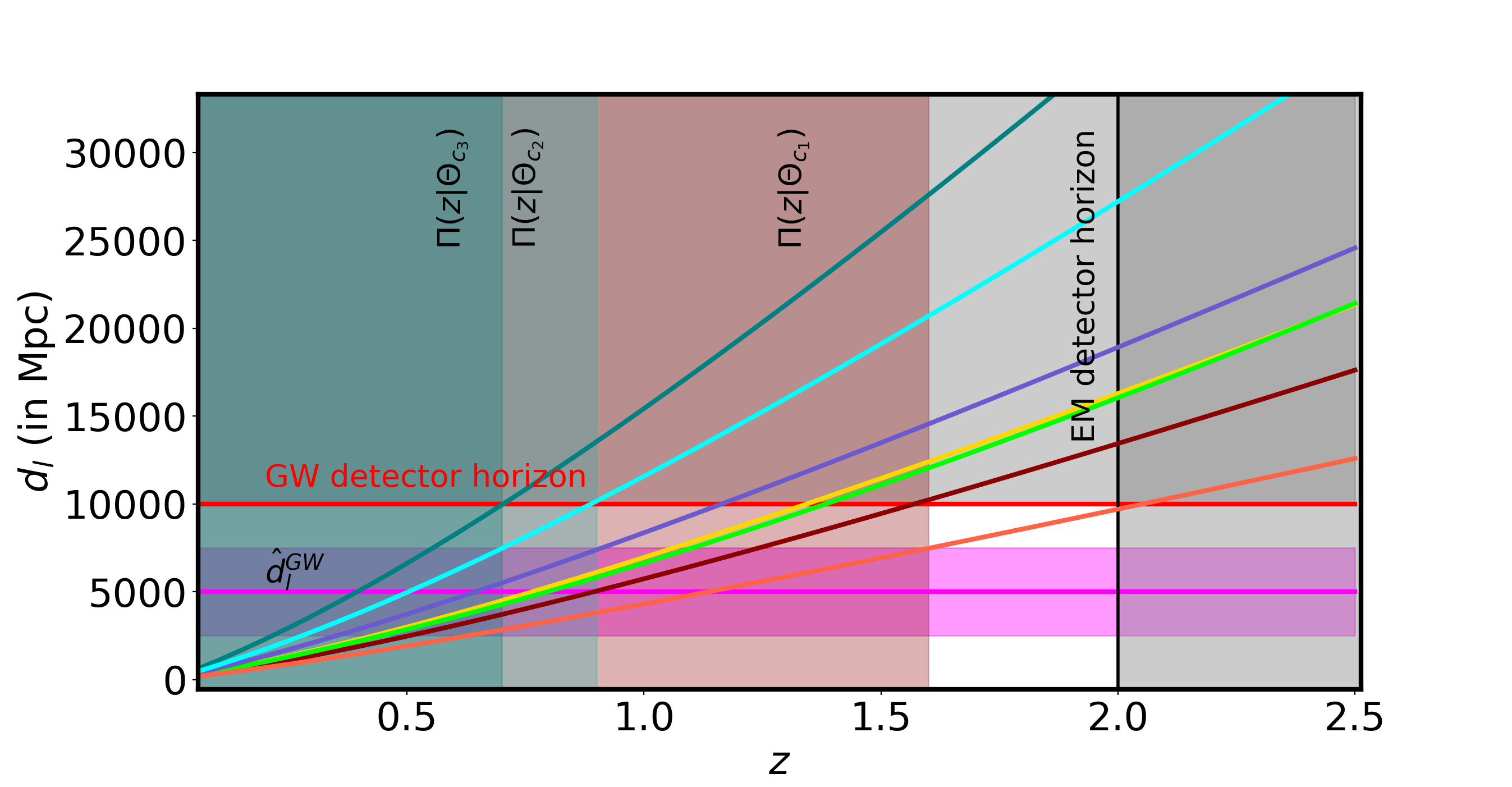}
\captionsetup{singlelinecheck=on,justification=raggedright}
\caption{Schematic diagrams showing the two scenarios  with EM counterpart (top) and without EM counterpart (bottom). We plot the luminosity distance $d_l$ and redshift $z$ plane for different cosmological models by different colors. The line in red shows the GW horizon in luminosity distance $d_{max}$ for a given GW detector noise and GW parameters in the detector-frame. For this schematic diagram, we have assumed $d_{max}= 10\,\textrm{Gpc}$. The line in black indicates the horizon for an EM follow-up instrument, which is assumed to be redshift $z=2$. The region shaded in grey above and right-side of the red and black line respectively are inaccessible to the GW and EM detectors. The region in magenta shows the luminosity distance $\hat d_l^{GW}$ inferred from a GW observation with the corresponding $1-\sigma$ error bar by the shaded region. In the top plot, we show the corresponding redshift $\hat z$ identified from EM follow-up in blue. In the bottom panel, we show the priors $\Pi(z, \Theta_{c_i})$ on redshift one needs to choose in the absence of an EM counterpart for different choices of the cosmological parameters $\Theta_{c_i}$. Different shaded regions in  teal, cyan and red color extend from $z=0$ to $z_{max}= z (d_{max}, \Theta_{c_i})$ which corresponds to the different parameter choices $\Theta_{c_i}$ (the corresponding distances are plotted in the same colors). The change in the maximum allowed redshift values with the change in cosmological parameters, for a fixed $d_{max}$ makes the priors  depend on cosmology.}
\label{fig:w}
\end{figure*}

We will explain this aspect in more details with a schematic diagram given in Fig. \ref{fig:w} which shows the luminosity distance $d_l$ and redshift $z$ plane for the case with EM counterpart (top) and case without EM counterpart (bottom). In the presence of an EM counterpart, one has a measurement of the redshift $\hat z$ (shown in blue) and the corresponding luminosity distance $\hat d_l^{GW}$ (shown in magenta). The combination of both these leads to estimate the best-fit cosmological parameters by fitting the luminosity distance and redshift relation, which are plotted for different choices cosmological parameters in different colors. The horizons for GW and EM observations which are shown by the red line and black line sets the maximum luminosity distance $d_{max}$ and maximum redshift $z_{max}$ accessible from GW detectors and EM detectors respectively\footnote{EM observations will also have a cutoff in luminosity distance. But for this schematic diagram, we have assumed that it is much larger than $d_{max}$.}. So, the region not shaded in grey is the total accessible region in the luminosity distance redshift plane. 

In the bottom panel we show the case when there is no EM counterpart. Here, one needs to choose a prior on the redshift $\Pi(z,\Theta_{c_i})$. Even for a fixed value of $d_{max}$, this prior depends on the cosmological parameters $\Theta_{c_i}$ as shown by the shaded regions in teal, cyan and red which corresponds to the maximum redshift $z_{max}= z(d_{max}, \Theta_{c_i})$ (for any cosmological model (shown in different colors)  $z_{max}$ is the redshift where the luminosity distance redshift curve   intersects the maximum luminosity distance $d_{max}$). So, depending on the choice of the cosmological parameters the range of allowed redshift varies. The total accessible parameter space is a combination of the allowed prior on redshift $z$ and luminosity distance $d_{max}$, which is also cosmology dependent. As a result, it is important to also include the probability associated with the population of the GW sources and its merger rates to the corresponding redshifts which are allowed by the prior. This is because for certain choices of the cosmological parameters, the value of $z_{max}$ can be large enough that the GW sources of stellar-origin are unlikely to be produced. However, when EM counterpart is present, such choices regarding the population are not required, as the association of a pair of GW and EM signal and the corresponding association of the  luminosity distance $\hat d_l^{GW}$ and redshift $\hat z$ pair are made using the time-domain information under the assumption of an astrophysical model $I_A$, as shown in Eq. \ref{eq:em}. Under an extremely rare scenario, if one identifies two EM counterpart originating from two different redshift within the same sky patch of the GW signal. Then one can use the population-based model to associate higher probability of being the EM counterpart to one of the events over the other.  

The EM counterpart gives a measurement of the redshift $\hat z$, and sky directions $\hat \theta, \hat \phi$. We assume that the EM data is accurately known. By using the luminosity distance $d_l(z, \Theta_c)$ and redshift $z$ relation, we can obtain the cosmological parameters $\Theta_c$ using the Bayes theorem, which can be written as 
\begin{eqnarray}\label{eq:gw}
&\mathcal{P}(\Theta_c|d_{GW}, \vec d_{EM}, I_A) \equiv \mathcal{P}(\Theta_c|d_{GW}, \vec d_{EM}) \propto  \frac{\Pi(\Theta_c)}{\beta(\Theta_c)} \\ \nonumber  & \times \int d\vec \Theta_{GW} \int dd_l  P(d_{GW}|d_l,\Theta_c, \vec d_{EM}, \vec \Theta_{GW})  P(\vec \Theta_{GW}|\vec d_{EM})\\\nonumber & \times P(d_l),
\end{eqnarray}
where $\Pi(\Theta_c)$ is the prior on the cosmological parameters $\Theta_c \in \{H_0, \Omega_m, w_0, w_a\}$, $P(d_{GW}|d_l,\Theta_c, \vec d_{EM}, \vec \Theta_{GW})$ is the likelihood given the EM data, $P(\vec \Theta_{GW}|\vec d_{EM})$ is the probability distribution of the GW source parameters $\vec\Theta_{GW} \in\{M_z, i, \Omega_{sky}\}$ (such as detector frame mass $M_z$, inclination angle $i$, sky solid angle $\Omega_{sky}$), given the EM data set. This is useful in converting the detector-frame mass $M_z$ to the source-frame mass $M= M_z/(1+\hat z)$, understanding about the inclination angle $i$ using the EM observation such as the astrophysical jet \citep{Mooley:2018dlz, 2018ApJ...868L..11M}, and identifying the sky localization of the GW source using the sky direction of the EM counterpart.  $P(d_l)$ is the prior on the luminosity distance. After marginalizing over the GW source parameters $\Theta_{GW}$, we can simplify Eq. \ref{eq:gw} as
   \begin{eqnarray}\label{eq:gw2}
    \mathcal{P}(\Theta_c|d_{GW}, \vec d_{EM}) 
    \propto \frac{\Pi(\Theta_c)}{\beta(\Theta_c)} & \int dd_l P(d_{GW}|d_l, \vec d_{EM}, \Theta_c) P(d_l),\,\,\,\,\,\,
\end{eqnarray}
where $d_{GW}$ is the luminosity distance marginalised over all the GW source parameters for the fixed redshift $\hat z$, and sky direction $\hat \theta, \hat \phi$ available from $\vec d_{EM}$. 
The normalization factor $\beta(\Theta_c)$  can be written as
\begin{eqnarray}\label{eq:beta1}
    \beta (\Theta_c)=& \int d\vec \Theta_{GW} \int d_{GW} \int \vec d_{EM} \int dd_l P(d_{GW}|d_l, \theta, \phi, \vec \Theta_{GW})\,\,\,\,\,\\\nonumber & \times P(\vec \Theta_{GW}|\vec d_{EM})P(d_l).
\end{eqnarray}
Assuming the EM counterparts are detected up to a maximum redshift $z_{max}$ with isotropic sensitivity in all directions and similarly, the GW sources are detected up to a maximum  luminosity distance $d_{max} (\theta, \phi, \vec \Theta_{GW})$. So, we can write $P(d_{GW}|d_l(\hat z, \Theta_c), \theta, \phi, \vec \Theta_{GW})= \mathcal{H}(d_{max}(\theta, \phi, \vec \Theta_{GW})-d_l)$\footnote{$\mathcal{H}(x)$ is the Heaviside step function.}. Then Eq. \ref{eq:beta1} becomes   
\begin{eqnarray}\label{eqbeta2}
    \beta (\Theta_c)=& \int d\vec \Theta_{GW} \int dd_l \int d_{GW}  \int \vec d_{EM}  P(\vec \Theta_{GW}|\vec d_{EM})\\\nonumber & \times \mathcal{H}(d_{max}(\theta, \phi, \vec \Theta_{GW})-d_l)P(d_l),\\\nonumber
    =& \int d\vec \Theta_{GW} \int_0^{d_{max}(\theta, \phi, \vec \Theta_{GW})} dd_l  P(\vec \Theta_{GW})P(d_l),\\ \nonumber =& \textrm{constant}.
\end{eqnarray}
The above integration is independent of the cosmological parameters and depends only on the maximum value of the luminosity distance $d_{max}$ and maximum redshift $z_{max}$, similar to the conclusion obtained by previous analysis \citep{Abbott:2017xzu,farr}. So, we can ignore this in the overall normalization in the Eq. \ref{eq:gw2}. In the Bayesian formalism mentioned in  Eq. \ref{eq:gw}, we have not included the correction due to the peculiar velocity, as it is not relevant for the source GW190521. However, previous studies have elaborately discussed the Bayesian formalism for the peculiar velocity correction of GW sources \citep{Abbott:2017xzu, Feeney:2018mkj, Mortlock:2018azx, Mukherjee:2019qmm, Nicolaou:2019cip} which can be easily incorporated in Eq. \ref{eq:gw}. 
\end{appendix}
\end{document}